\documentclass[aps,prb,reprint]{revtex4-1}
\usepackage{graphicx}
\usepackage{dcolumn}
\usepackage{bm}
\usepackage{amssymb}
\usepackage{amsmath}
\usepackage{epsf}
\usepackage{color}
\draft

\begin{document}
\preprint{APS/123-QED}

\title{Field-cooled state of the canonical spin glass revisited}

\author{Sudip Pal, Kranti Kumar, A. Banerjee, S. B. Roy}
\email{sbroy@csr.res.in}
 \affiliation{%
 UGC DAE Consortium for Scientific Research\\
 Indore-452001, India 
}%

\author{A. K. Nigam}
\affiliation{%
 Tata Institute of Fundamental Research\\
 Mumbai-400005, India 
}%
\date{\today}
\begin{abstract}
Canonical spin glass (SG) is an enigmatic system in condensed matter physics. In spite of the intense activities of last five decades several questions regarding the nature of the SG phase transition and the SG ground state are yet to be resolved completely. In this backdrop we have revisited the field cooled state of canonical spin glass. We have experimentally studied magnetic response in two canonical spin-glass systems AuMn(1.8\%) and AgMn(1.1\%), both in the field cooled (FC) as well as zero field cooled (ZFC) state. We show that the well known magnetic memory effect, which clearly established earlier the metastable nature of the ZFC state in SG, is also present in the FC state. The results of our experimental study indicate that the FC state also is a non-equilibrium state, contrary to the common understanding and hence the energy landscape involved is a non-trivial one. This in turn seriously questions the picture of spin-glass transformation as a simple second order thermodynamic phase transition.
\end{abstract}
                             
\maketitle

Spin-glass (SG) transition in magnetic alloy systems is a very interesting phenomenon in condensed matter physics {\color{blue}\cite{RMP1986, Mydosh2015}} and the concept has found applications in various other areas like colloids, granular media, structural glass, neural networks etc.{\color{blue}\cite{Granular1992, Sherrington1993,Parisi2006}} Over the years there have been two distinct approaches to understand this SG transition in canonical spin glass systems. First, the Parisi's solution of  infinite range Sherrington-Kirkpatrick (SK) model that predicts a thermodynamic second order phase transition from high temperature paramagnetic state to low temperature spin-glass state.{\color{blue}\cite{SK1975, Parsi1983, DC1984}} Within this model cooling the system in presence of a finite field from paramagnetic state would bring the spin glass to the ``infinite time equilibrium state".{\color{blue}\cite{Mydosh2015}} The other model, known as a short range droplet model, predicts a SG transition only at zero magnetic field (H) and absence of any phase transition for non-zero H in any finite dimension.{\color{blue}\cite{Young1, Fisher1986, Moore1998}} Numerical simulations have found a SG phase transition in presence of  external (random) field only above an upper critical dimension of nearly six{\color{blue}\cite{Young2,Helmut2009}}; the debate, however, is yet to be resolved completely.{\color{blue}\cite{Helmut2012,Michele2015}}

There exist some experimental reports showing that the field cooled (FC) magnetization in spin-glasses being sensitive to the cooling rate and the time decay of thermoremnant magnetization depended on the waiting time before the cooling field was removed.{\color{blue}\cite{Wang, Chamberlin1984, Bouchiat1985, Nordblad1986}} This contradictis the picture of a simple second order phase transition to an equilibrium state. However, the FC state of spin glass has not been a subject of intense scrutiny as much as the zero field cooled (ZFC) state so far, presumably due to the popularity and the implicit acceptance of thermodynamic phase transition picture.{\color{blue}\cite{Granular1992, Sherrington1993,Parisi2006}} In this light it is necessary to revisit the FC state of the canonical spin-glasses and subject it to fresh experimental scrutiny. 

Memory effects observed in various glassy systems have emerged as effective experimental technique to characterize the energy landscape of such systems.{\color{blue}\cite{Samarakoon2016,MERMP}} A system in equilibrium state does not show memory effect, whereas a system with rugged energy landscape can store the memory of a previous state traversed while cooling, and it affects the physical properties of the system during the subsequent re-heating. The ZFC state of a canonical spin glass shows memory effect in dc magnetization measurement, thus clearly highlighting the metastable nature of this state. In the present work, we have investigated the memory effect in two well-known canonical spin-glass systems AuMn(1.8\%) and AgMn(1.1\%) through dc magnetization measurements. In particular, we have studied the nature of the FC state of these canonical SG systems through the observation of memory effects in low field dc magnetization. We find that along with the ZFC state, the FC state also shows clear signature of memory effect below the spin-glass freezing temperature. This highlights the non-equilibrium nature of the FC state in canonical spin glasses AuMn(1.8\%) and AgMn(1.1\%).

\begin{figure}[h]
\centering
\hspace{-0.5cm}
\includegraphics[scale=0.45]{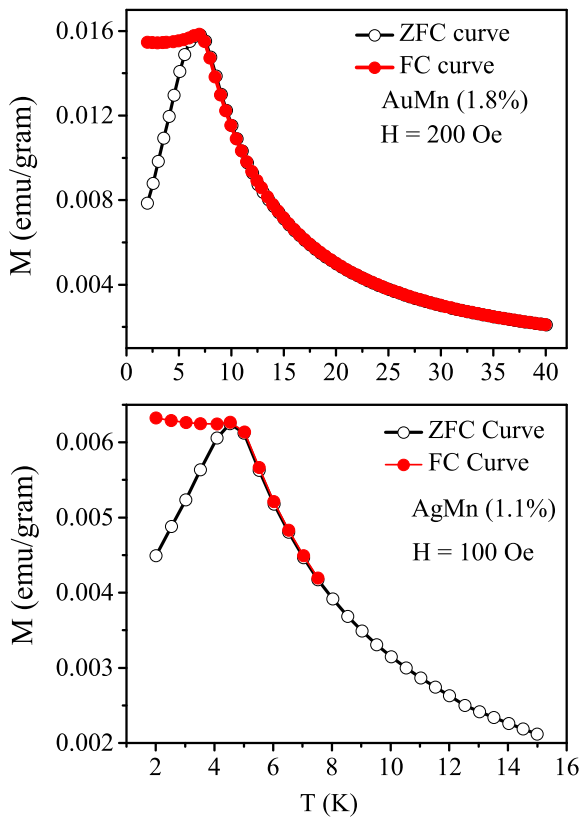}
\caption{Magnetization versus temperature plots for AuMn(1.8\%) and AgMn(1.1\%) in applied magnetic fields H = 200 and 100 Oe, respectively.}
\end{figure}

\begin{figure}[h]
\centering
\hspace{-0.5 cm}
\includegraphics[scale=0.32]{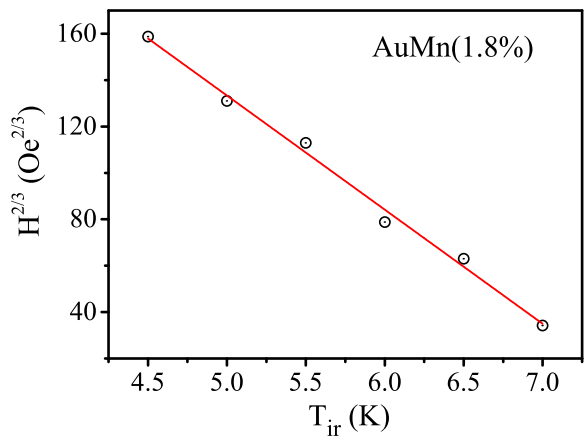}
\caption{Variation of irreversibility temperature T$_{ir}$ in H-T plane for AuMn(1.8\%). Red line is the linear fit to the data.}
\end{figure}

AuMn(1.8\%) and AgMn(1.1\%) samples have been prepared by induction melting in Argon atmosphere. Further details of the sample preparation and characterization have been reported earlier.{\color{blue}\cite{Nigam1983}} Magnetization measurements have been carried out in 7 Tesla SQUID magnetometer (M/S Quantum design, USA) using Reciprocating Sample Option (RSO) transport with scan length of 4 cm both in the ZFC and FC mode. In the ZFC mode the sample is cooled to the lowest temperature (here 2 K) before the applied magnetic field (H) is switched on, and the measurement is made while warming up the sample. In the FC mode the sample is cooled to the lowest temperature in the presence of applied H and the measurement is then made while warming up the sample. All measurements have been performed at a fixed cooling and heating rate of 0.2 K/min. 

\begin{figure}[h]
\centering
\includegraphics[scale=0.32]{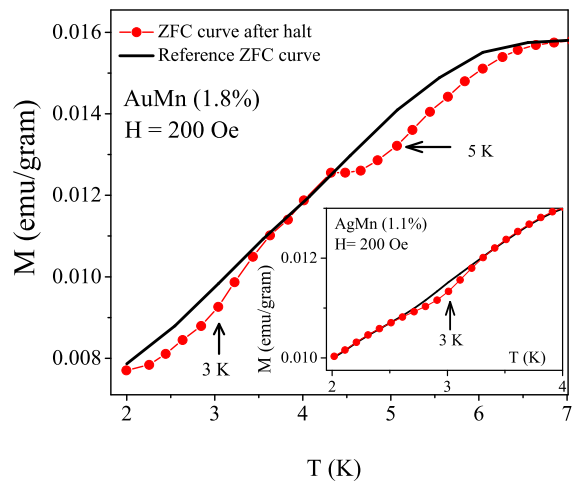}
\caption{ Zero-filed-cooled (ZFC) magnetization memory effect in AuMn(1.8\%) at H$_m$ = 200 Oe. In one experiment, the magnetization was measured under a standard ZFC protocol (see text for details). In the other, cooling was stopped and temperature kept constant at T = 3 and 5 K for 3 h while cooling in zero field. The inset shows the data for AgMn(1.1\%) where cooling was stopped at T= 3 K for 3 h.}
\end{figure}

\begin{figure}[h]
\centering
\includegraphics[scale=0.34]{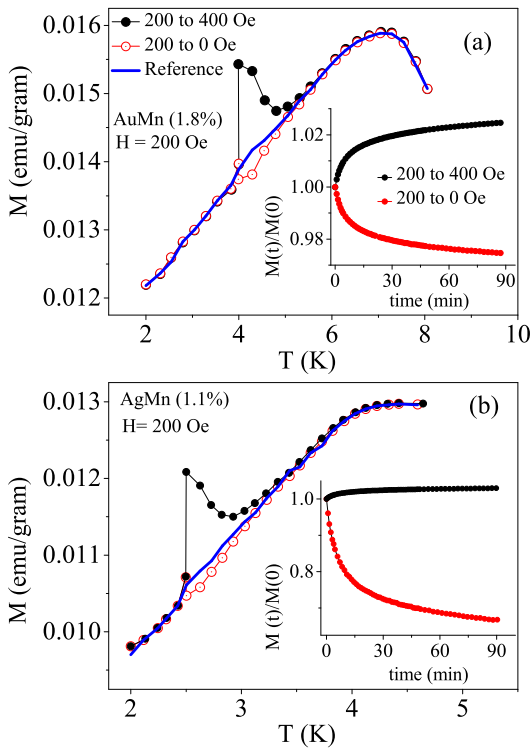}
\caption{ (a) ZFC memory effect in AuMn(1.8\%) at H$_m$ = 200 Oe following two separate measurement protocols. While warming after zero field cooling, at T$_m$ = 4 K, in first protocol, H is increased to 400 Oe (black solid circles) and in the second protocol, H is reduced to zero (red dotted circles) and held for t$_w$ = 1.5 h. The field is then returned back to 200 Oe and heating is continued. Inset shows the aging of the spin glass during waiting time. (b) shows the results for AgMn (1.1\%) under a similar protocol at T$_m$=  2.5 K. }
\end{figure}

\begin{figure}[t]
\centering
\includegraphics[scale=0.4]{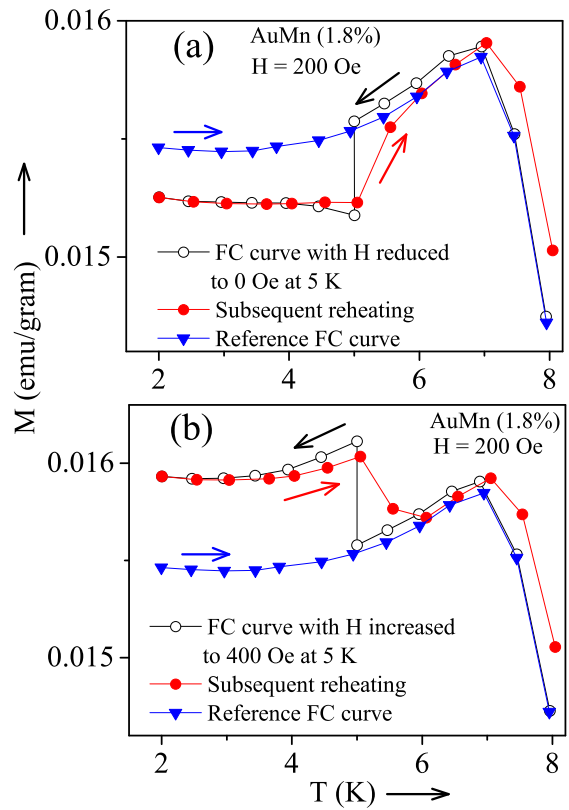}
\caption{ Field cooled (FC) magnetization memory effect in AuMn(1.8\%) at H$_m$ = 200 Oe and  T$_m$ = 5 K. (a) field is reduced to 0 Oe (remnant)  for a time period of t$_w$ = 1.5 hour and returned back to H$_m$ = 200 Oe. (b) field is increased from 200 to 400 Oe. Then the system is subsequently cooled and re-heated.}
\end{figure}

AuMn(1.8\%) and AgMn(1.1\%) undergo the spin-glass or spin freezing transition around T$_f$ = 7 and 5 K respectively as evident from the cusp in the ZFC magnetization curve as shown in Fig. {\color{blue}1}. The T$_f$s obtained in the present study agrees quite well with the earlier reported results.{\color{blue}\cite{Nigam1983}} The ZFC and FC  M(T) curves bifurcate below an irreversibility temperature, T$_{ir}$ which lies slightly below  T$_f$.  As we increase the applied H (data not shown here), both T$_{ir}$ and T$_f$ gradually shift towards lower temperatures and the difference between FC and ZFC magnetization simultaneously reduces. The cusp at T$_f$ in the ZFC curve also broadens with applied H. In Fig. {\color{blue}2}, we have shown the variation of  T$_{ir}$ with respect to H for the sample AuMn(1.8\%). It approximately varies according to T$_{ir}$ $\sim$ H$^{2/3}$ showing the existence of  de Almeida - Thouless (AT) line in the H-T plane as predicted by the mean field theory of spin glass.{\color{blue}\cite{ATline}} 

In the main panel of Fig. {\color{blue}3}, we present the memory effect observed in AuMn(1.8\%) following standard ZFC procedure. Here we have first cooled the sample in absence of any magnetic field down to 2 K and then measured the magnetization of the sample while heating in the presence of an applied H of 200 Oe. This provides the reference ZFC magnetization curve. The sample is then cooled again in zero magnetic field but with temporary halts at T$_m$ = 5 and 3 K for 3 hours at each temperature. The measurement of magnetization is then performed as before while heating. The magnetization shows distinct dip around each T$_m$ before merging with the reference ZFC curve at higher temperatures (see Fig. {\color{blue}3}). Thus it reflects the strong memory of cooling history and highlights the metastable nature of the ZFC state of the spin-glass. AgMn(1.1\%) also shows similar behavior when measured following similar protocol with T$_m$ = 3 K and H= 200 Oe (see inset of Fig. {\color{blue} 3}).  A simple trap model of complex distribution of multiple potential minima in the free energy landscape has been widely used to explain these features.{\color{blue}\cite{Mathieu1, Mathieu2, Bouchaud1992, Bouchaud1995, Vincent2009}} According to this model, potential minima are separated by finite energy barriers and the system is trapped in the minima for certain trapping time. The minima get hierarchically divided into more number of minima as temperature decreases. Simultaneously, the barriers become more effective due to reduced thermal energy, and the heights of some of the barriers also increase.{\color{blue}\cite{Vincent, Hamman}} Therefore, trapping time increases with reduction of temperature. During the waiting time, the system gradually creeps through the lower energy states available nearby. This is known as ``aging''.  When cooling is resumed after aging, the system ``rejuvenates'' back to the earlier state and magnetization shows a finite dip in the following reheating curve (see Fig. {\color{blue}3}). It is further reported that the application of a small magnetic field for finite time while zero field cooling, also results in memory effect.{\color{blue}\cite{Schmitt2013}} This indicates that a SG also remembers the evolution of its state due to external magnetic field. We have used here a different protocol to find the effect of field cycle on the ZFC memory, which is shown in fig. {\color{blue}4 (a)} and {\color{blue}4 (b)} for AuMn(1.8\%) and AgMn (1.1\%) respectively. In this protocol, we have initially cooled the system in zero field down to 2 K, applied the measuring field H$_m$ = 200 Oe and then start recording the ZFC curve. At an intermediate temperature T$_m$ below T$_f$ (T$_m$= 4 and 2.5 K for  AuMn(1.8\%) and AgMn (1.1\%) respectively), cooling was temporarily paused and the field was increased to 400 Oe for next t$_w$ = 1.5 hour and again reduced back to 200 Oe and warming was resumed. After the field cycle, magnetization starts from a higher value and merges back to the reference ZFC curve on further heating (see black solid circle in the main panel of Fig. {\color{blue}4 (a), (b)}).  Magnetization increased with time during the wait time t$_w$, and this increase could be fitted with stretched exponential function (see the insets of fig. {\color{blue}4(a), (b)}, black curve). On the other hand, in another measurement where the field is reduced from H$_m$ = 200 Oe to zero at T$_m$ instead of increasing it, magnetization starts form a lower value, however, very small (as shown in the Fig. {\color{blue}4}, red dotted circle,) and merges back with the reference curve. These measurements clearly show that a spin glass remembers the state that is perturbed by an external field. Most interestingly, in case of AuMn(1.8\%), both the increase and decrease in magnetic field perturb the state by nearly same amount (evident from amount of relaxation, inset of Fig. {\color{blue}4(a)}). On the contrary, in case of AgMn(1.1\%), relaxation is significantly larger when H is reduced compared to the field increasing cycle. However, the increase in field causes the larger discontinuity in both systems. Therefore, effect of field cycle is highly assymetric and independent of relaxation. We note here that we have also performed these measurements in different temperatures below T$_f$ and the assymetric response of the ZFC cooled state under field cycle is evident in all measurements, which require further investigations.    

Fig. {\color{blue}5} presents the memory effect in the FC state of the AuMn(1.8\%) sample. Here, we have first measured the reference FC magnetization curve at applied field, H$_m$ = 200 Oe. To investigate the memory effect, we have followed two protocols. We cool the system from the paramagnetic state to T$_m$, which is less than the respective T$_{ir}$, the irreversibility temperature of the sample for the applied field H$_m$. Then, in our first protocol, we have reduced the field to zero and after a waiting period of t$_w$ = 1.5 hour, the field H$_m$ is reapplied and cooling of the sample is resumed taking it down to 2 K and then subsequently reheated. In the second protocol, at T$_m$ instead of reducing the field to zero, it is increased to 400 Oe and kept fixed for next 1.5 hour and then reduced back to H$_m$ = 200 Oe. Then the cooling in presence of 200 Oe is resumed and subsequently re-heated. In both protocols we see a discontinuity in the cooling curve due to the field cycle. However, in the first protocol, magnetization after the field cycle (and wait for 1.5 hour) is smaller than the M value before performing the cycle. The subsequent warming curve also shows discontinuity above T$_m$ and finally merges with the reference warming curve showing the memory effect. In the second protocol, in which the field was increased from H$_m$= 200 to 400 Oe, M again shows a discontinuity but in this case, M after the field cycle is larger than the M before the cycle. Here also the consequent warming M-T curve shows memory effect and merges with the reference M-T curve. Fig. {\color{blue}6} shows the data for AgMn(1.1\%) following similar measurement protocol while performing the field cycle at T$_m$ = 3 K.
  
\begin{figure}[t]
\centering
\includegraphics[scale=0.4]{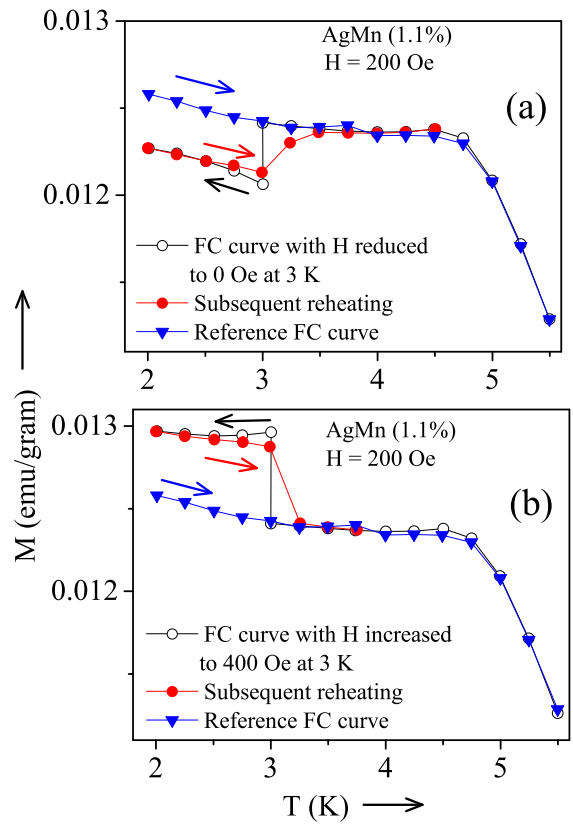}
\caption{ Field cooled (FC) magnetization memory effect in AgMn(1.1\%) at H$_m$ = 200 Oe at T$_m$ = 3 K. (a) field is reduced to 0 Oe (remnant) for a time period of t$_w$ = 1.5 hour and returned back to H$_m$ = 200 Oe. (b) field is increased from 200 to 400 Oe. Then the system is subsequently cooled and re-heated.}
\end{figure}

The above results clearly indicate that even in the FC state a spin-glass remembers the earlier history. According to the short range interaction model of SG {\color{blue}\cite{Young2}}, FC state is a paramagnetic state. Whereas, in the Mean field model {\color{blue}\cite{Mydosh2015}} the field cooling brings the system to an equilibrium state. In the later case the system is not supposed to have any memory effect, for example a long range ferromagnetic state reached from a high T paramagnetic state through a second order phase transition does not show any memory effect although it has degenerate states in absence of any symmetry breaking field. Therefore, the memory effect in the FC state of spin glass possibly hints towards a non-trivial nature of the energy landscape of the system in it's field cooled state as well, which cannot be reconciled within the framework of both the infinite range SK model {\color{blue}\cite{SK1975, Parsi1983, DC1984}} and droplet model {\color{blue}\cite{Young1, Fisher1986, Moore1998}} of the spin glass. It may be noted here that there exist some early experimental results involving measurements of time relaxation of thermoremanent magnetization in some canonical spin-glasses (like AgMn(2.6$\%$), CuMn(5$\%$) with relatively higher concentration of transition metals), which were suggestive of the fact that the FC state of spin glass may also have considerable dynamics.{\color{blue}\cite{Wang, Chamberlin1984, Bouchiat1985, Nordblad1986}} In these measurements, both the ZFC and FC state of spin glass were found to show similar aging dynamics depending on the waiting time given to the system before the field was changed. Indeed it was suggested  that even though the FC magnetization was constant in time, the FC state was not in equilibrium for finite wait time $\it t_w$. {\color{blue}\cite{Chamberlin1984}} Our present results further reinforce those earlier observations. 

The memory effect in the FC state has been reported earlier in many glassy magnetic systems like, cluster glass, assembly of magnetic nanoparticles etc.{\color{blue}\cite{Sun2003}} In these systems, the memory effect arises from the wide distribution of relaxation time originating because of competing interaction between magnetic clusters as well as distribution in cluster size.{\color{blue}\cite{Sun2003, Sasaki2005, Chakraverty2005}} It may be noted here that apart from the two main stream theoretical models namely SK model and droplet model, a cluster model of spin glass has also been considered for canonical spin glass systems.{\color{blue}\cite{Klein1963,Smith1974, Levin1979}} According to this model, the long range interactions between spins may encourage formation of the correlated spin clusters even above the spin freezing temperature. These clusters form at relatively higher temperature primarily due to concentration fluctuation of impurity spins and they grow in size with decreasing temperature. Below the characteristic temperature, T$_f$, an infinite cluster forms along with many other finite clusters with varying size and shape. Within one cluster each spin is connected with at least one of the other spins in the same cluster. This kind of system has broad distribution of relaxation spectrum and therefore can possibly be useful in explaining the observed memory effects. 

In conclusion, we have studied memory effect in the ZFC and FC state of two canonical spin glass systems AuMn(1.8\%) and AgMn(1.1\%) through dc magnetization measurements. We find that ZFC state shows clear memory effect which is reminiscent of the rugged free energy landscape of spin glass. In addition, distinct memory effect is also observed in the FC state. Change in applied magnetic field perturb the FC state at a temperature T$_m$ below T$_{ir}$. In the subsequent re-heating cycle, the system returns to the earlier state above T$_m$, similar to the ZFC state, signifying that the FC state of a canonical spin glass also contains a rugged energy landscape, which is not consistent with the picture of a thermodynamic equilibrium state. This possibly indicates that the theoretical understanding of spin-glass phenomena is yet to be complete, and the cluster model {\color{blue}\cite{Klein1963,Smith1974, Levin1979}} for spin-glass may need some reconsideration for accounting such experimental observations of metastable behavior.


\begin{thebibliography}{Bibliography}
\bibitem{RMP1986} K. Binder and A. P. Young, Rev. Mod. Phys. {\bf 58} 801 (1986).
\bibitem{Mydosh2015} J. A. Mydosh, Rep. Prog. Phys. {\bf 78}  052501 (2015).
\bibitem{Granular1992}H. M. Jaeger, Sidney R. Nagel, Science, {\bf 255}, 1523 (1992).
\bibitem{Sherrington1993} D. Sherrington, in Mathematical approaches to Neural Networks", ed.
J.G. Taylor (Elsevier), p261 (1993).
\bibitem{Parisi2006} G. Parisi, PNAS {\bf 103} 7948 (2006).
\bibitem{SK1975} D. Sherrington and S. Kirkpatrick, Phy. Rev. Lett. {\bf 35}, 1792 (1975).
\bibitem{Parsi1983} G. Parisi, Phys. Rev. Lett. {\bf 50}, 1946 (1983).
\bibitem{DC1984} D. Choudhury and A. Mookerjee, Phys. Rep. {\bf 114} 1 (1984).
\bibitem{Young1} A. P. Young, J. Phys. A: Math. Theor. {\bf 41}, 324016 (2008); J. Mattsson, T. Jonsson, P. Nordblad, H. Aruga Katori, A. Ito,  Phys. Rev. Lett. {\bf 74}, 4305 (1995).
\bibitem{Fisher1986} D. S. Fisher and D. A. Huse, Phys. Rev. Lett. {\bf 56}, 1601(1986).
\bibitem{Moore1998} M. A. Moore, Hemant Bokil, and Barbara Drossel, Phys. Rev. Lett. {\bf 81}, 4252 (1998).
\bibitem{Young2} A. P. Young and H. G. Katzgraber, Phys. Rev. Lett. {\bf 93}, 207203 (2004).
\bibitem{Helmut2009} H. G. Katzgraber, D. Larson and A. P. Young, Phys. Rev. Lett. {\bf 102}, 177205 (2009).
\bibitem{Helmut2012} H. G. Katzgraber, Thomas Jorg, Florent Krzakała, and Alexander K. Hartmann, Phys. Rev. B {\bf 86}, 184405 (2012).
\bibitem{Michele2015} M. Castellana and C. Barbieri, Phys. Rev. B {\bf 91}, 024202 (2015).
\bibitem{Wang} Tie Wang, H. V. Bohm and L. E. Wenger, J. Magn. Magn. Mat. {\bf 54-57}, 89 (1986).
\bibitem{Chamberlin1984} R. V. Chamberlin, Phys. Rev. B {\bf 30}, 5393 (1984).
\bibitem{Bouchiat1985} H. Bouchiat and D. Mailly, J. Appl. Phys. {\bf 57}, 3453 (1985).
\bibitem{Nordblad1986} P. Nordblad, L. Lundgren and L. Sandlund, J. Magn. Magn. Mat. {\bf 54-57}  185 (1986) .
\bibitem{Samarakoon2016} Anjana Samarakoona,  Taku J. Satob, Tianran Chena, Gai-Wei Cherna, Junjie Yanga, Israel Klicha, Ryan Sinclairc, Haidong Zhouc, and Seung-Hun Lee, PNAS {\bf 113}, 11806 (2016).
\bibitem{MERMP} N. C. Keim, Joseph D. Paulsen, Zorana Zeravcic, Srikanth Sastry, Sidney R. Nagel, Rev. Mod. Phys. {\bf 91}, 035002 (2019).
\bibitem{Nigam1983} A. K. Nigam and A. K. Majumdar, Phys. Rev. B {\bf 27}, 495 (1983).
\bibitem{ATline} J R L de Almeida and D. J. Thouless, J. Phys. A: Math. Gen. {\bf 11}, 983 (1977).
\bibitem{Mathieu1} R. Mathieu, P. J$\ddot{o}$nsson, D. N. H. Nam, and P. Nordblad, Phys. Rev. B {\bf 63}, 092401 (2001). 
\bibitem{Mathieu2} R. Mathieu, H. Hudl and P. Nordblad, Euro Phys. Lett. {\bf 90}  67003 (2010).
\bibitem{Bouchaud1992} J. P. Bouchaud, J. Phys. I France {\bf 2} 1705 (1992).
\bibitem{Bouchaud1995} J.P. Bouchaud and D.S. Dean, J. Phys. I France {\bf 5} 265 (1995).
\bibitem{Vincent2009} E. Vincent, J. Hammann and M. Ocio, J Stat Phys  {\bf 135} 1105 (2009).
\bibitem{Vincent} E. Vincent, J. P. Bouchaud, J. Hammann, and F. Lefloch Philos. Mag. {\bf 71},  489 (1995).
\bibitem{Hamman} J. Hammann, M. Lederman, M. Ocio, R. Orbach and E. Vincent, Physica A {\bf 185}, 278 (1992).
\bibitem{Schmitt2013} Devin C. Schmitt, Joseph C. Prestigiacomo, Philip W. Adams, David P. Young, Shane Stadler, and Julia Y. Chan, App. Phys. Lett. {\bf 103}, 082403 (2013).
\bibitem{Sun2003}Young Sun, M. B. Salamon, K. Garnier and R. S. Averback, Phys. Rev. Lett. {\bf 91}, 167206 (2003).
\bibitem{Sasaki2005}M. Sasaki, P. E. J$\ddot{o}$nsson, H. Takayama and H. Mamiya, Phys. Rev. B {\bf 71}, 104405 (2005).
\bibitem{Chakraverty2005} S. Chakraverty, M. Bandyopadhyay, S. Chatterjee, S. Dattagupta, A. Frydman, S. Sengupta, and P. A. Sreeram, Phys. Rev. B {\bf 71}, 054401 (2005).
\bibitem{Klein1963} M. W. Klein and R. Brout, Phys. Rev. {\bf 132}, 2413 (1963).
\bibitem{Smith1974} D. A. Smith, J. Phys. F: Met. Phys {\bf 4}, L266 (1974).
\bibitem{Levin1979} K. Levin, C. M. Soukoulis, and G. S. Grest,  J. Appl. Phys. {\bf 50}, 1695 (1979).
\end{thebibliography}
\end{document}